\documentclass[12pt,a4paper]{article}
\begin{document}
\large

\newpage
\begin{center}
{\bf ELECTROMAGNETIC CHARACTERISTICS OF STATES $^{172}Yb$}
\end{center}
\vspace{0.5cm}

{\bf Ph.N. Usmanov$^{1,2}$,
A.A. Okhunov$^{}$\footnote{\emph{E-mail:}
aaokhunov@rambler.ru}, U.S. Salikhbaev$^{1}$,
R. Rajabbaev$^{3}$}
\vspace{0.5cm}

{\bf $^{1}$Institute of Nuclear physics Academy of Science of Uzbekistan, 702132,Tashken;\\
$^{2}$The Namangan engineering - economic institute, Namangan, Uzbekistan; \\
$^{3}$Specialization of high and middle educations development center
of MHMSE RUz, Tashkent, Uzbekistan.}
\vspace*{1.0cm}

\begin{abstract}
To describe the ground (gr) and excited states of even-even deformed nuclei
with the phenomenological model, which takes into account the mixing ground
$(gr)$, $0^{+}_{2}(\beta_1)$--, $0^{+}_{3}(\beta_2)$--, $2^{+}_{1}(\gamma_1)$--,
$2^{+}_{2}(\gamma_2)$-- and $K^{\pi}=1^{+}_{\nu}$ rotational bands the calculation
has been done for the isotope $^{172}Yb$. The energy specterum is described.
The reduced probability of electric quadrupole, monopole and dipole magnetic
transitions from the states $0^{+}_{\nu}$ and $2^{+}_{\nu}$ - bands to the
ground $(gr)$ state band is calculated. A non adiabatic effect in the energy
and electromagnetic characteristics is discussed.
\end{abstract}


\section{Introduction}

In last year the interest to study the properties of the deformed nuclei has especially
raised in connection with discovering of new collective isovector magnetic dipole modes.
The measured energy values under the mentioned above conditions of such fashions testify
that they are located not so high as the excitation spectrum, and the account of mixing
of conditions isovector magnetic modes with low - lying states can lead to interesting
physical phenomena \cite{Usmanov}.

The $^{172}Yb$ nucleus is the most studied nucleus. It has many clearly observed
$K^{\pi}=0^+$ bands with the developed rotational structure. Some of these bands
include levels with enough spins \cite{Bekzhanov}.
In $(n,\gamma)$ reactions, large number of $\gamma$-- transitions taking into
account conditions on levels below 2 $MeV$ are also investigated $\gamma$-- beams
unloading these conditions \cite{Bekzhanov}.

The experimental and theoretical works are devoted to studies of rotational levels
properties. In the majority of these works characteristics of electromagnetic transitions
\cite{Bekzhanov,Fahlander,Youhana,Usmanov} are studied.

Experimental values of $B(E2)$ on the low - lying levels $K^{\pi}=0^{+}_{\nu}$ and
$2^{+}_{\nu}$ bands \cite{Bekzhanov,Fahlander,Gresswell,Suber}, as well as ratio
values $X_I(E0/E2)$, dimensionless matrix elements $E0$-- transition $\rho(E0)$
and factor of multipole mixture $\delta(E2/M1)$ are available
\cite{Gresswell,Suber,Voinonova, Kracikova,Singh,Demidov}. Last years, were
dedicate to intensive experiments for definition of low - lying collective states
$K^{\pi}=1^+$ in the deformed
nuclei, in particular in isotopes $^{172,174,176}Yb$, and to development of their
theoretical researches.

The present work studies electromagnetic characteristics of the $^{172}Yb$ isotope. The
resulted probabilities and matrix elements for $E0$-- , $E2$-- and $M1$-- transitions,
as well as multipoles mixture factors are calculated. These calculations were performed
within the phenomenological model \cite{Usmanov}, in which mixing of low - lying
conditions, $\beta_{\nu}$, $\gamma_{\nu}$ and rotational bands are considered.

\section{Model. The energy spectrum}
The application of the phenomenological model gives more interesting information
about the states rotational bands Coriolis mixing to explain unexpected effects.
The basic states of the Hamiltonian include ground
$(gr)$, $0^{+}_{2}(\beta_1)$--, $0^{+}_{3}(\beta_2)$--, $2^{+}_{1}(\gamma_1)$--,
$2^{+}_{2}(\gamma_2)$-- vibrational and $K^{\pi}=1^{+}_{\nu}$ rotational bands.

The Hamiltonian of the model is written as \cite{Usmanov}:
\begin{eqnarray}
H&=&H_{rot}(I^2)+H_{K,K'}^{\sigma}(I)
\end{eqnarray}
here $H_{rot}(I^2)$-- is rotational part,
\begin{eqnarray}
H_{K,K'}^{\sigma}(I)=-\omega_K\delta_{K,K'}-\omega_{rot}(I)
(j_x)_{K,K'}\chi(I,K)\delta_{k,K'\pm1}
\end{eqnarray}
where $(j_x)_{K,K'}$-- is the matrix element (m.e.), describing the Coriolis
coupling of rotational bands, and $\omega_{rot}(I)$-- is the angular frequency
of core rotation $(\omega_{rot}(I)=dE_{cor}(I)/dI )$, $\omega_K$-- head
energy bands and
$$\chi(I,0)=1, \ \ \chi(I,1)=[1-\frac{2}{I(+1)}]^{1/2}$$

The eigen wave function of Hamiltonian is
\begin{eqnarray}
|IMK>&=&\sqrt{\frac{2I+1}{16\pi^2}}\left\{\sqrt{2}\Psi_{gr,K}^ID_{M,0}^I(\theta)
+\sum_{K}\frac{\Psi_{K',K}^I}{\sqrt{1+\delta_{K',0}}}\times\right. \nonumber\\
&&\left.\times\left[D_{M,0}^I(\theta)b_{K'}^+ + (-1)^{I+K'}D_{M,K'}^I(\theta)b_{-K'}\right]\right\}
\end{eqnarray}

here $\Psi_{K^{'},K}^I$ -- is the amplitudes of the basis state mixing. The states
consist of (5+$\nu$) bands, where $\nu$-- is the number of included $1^+$-- states.
It includes ground $|0>$ state bands and the single - phonon
$b_{\lambda-2,K}^+|0>=b_{K}^+|0>$ with $K^{\pi}=0_{\gamma}^+$, $2^+$ and
$1_{\nu}^+$-- rotational bands.

By solving the Schrodinger equation
\begin{eqnarray}
H_{K,n}^{\sigma}\Psi_{K,n}^I=\varepsilon_n^\sigma\Psi_{K,n}^I.
\end{eqnarray}

we obtained wave function and energy of states with positive parity.

Total energy of states is determined by following equation
\begin{eqnarray}
E_n^\sigma(I)=E_{rot}(I)+\varepsilon_n^\sigma(I).
\end{eqnarray}

Energy of rotational core $E_{rot}(I)$ can be determined by different methods,
for example, by Harris parameterization \cite{Harris} of the angular
moment and energy.
\begin{eqnarray}
\sqrt{I(I+1)}=J_0\omega_{rot}(I)+J_1\omega_{rot}^3(I) \nonumber\\
E_{rot}(I)=\frac{1}{2}_0\omega_{rot}^2(I)+\frac{3}{4}J_1\omega_{rot}^4(I)
\end{eqnarray}

\noindent
where $J_0$ and $J_1$-- parameters of inertia of rotational core. We propose
for small spin the energy of rotational core conceders with the energy of
ground rotational bands \cite{Bengtsson}. The values of these parameters were
obtained from equation (6) by using experimental data for the ground bands
energy for $I\leq 10$. Head energy ground band state and $\beta_n$-- band were
taken from experiments, so that they are not influenced by the Coriolis forces.
Matrix elements of Coriolis mixture rotational states $(j_x)_{K,K'}$ and head
energy $\gamma$-- bands ($\omega_{\gamma}$), were determined by fitting the
calculated energies spectera with positive parity states from experimental
data. In this case, $(j_x)_{K,1_1}=(j_x)_{K,1_2}$. Parameters of the
model are presented in Table 1. Theoretical and experimental spectra energy
are given in the figure 1. One can see from fig.1 that our model describes
the energy of positive parity states well. From the energy spectrum we can see
that energy $\beta_2$ and $\gamma_1$ bands are close to each other. These 
effects must be shown in the electromagnetic transitions, which is 
different their from
adiabatic values.
\begin{center}
\textbf{Table 1. \ \ \ Parameters of calculations }\\
\begin{tabular}{|c|c|c|c|c|c|c|c|c|} \hline
   \ \ $\omega_{\gamma_1}$ \ \ &  \ \ $\omega_{\gamma_2}$ \ \ &  \ \ $(j_x)_{gr,1_{\nu}}$ \ \ &
   \ \ $(j_x)_{\beta_1,1_{\nu}}$ \ \ &  \ \ $(j_x)_{\beta_2,1_{\nu}}$ \ \ &
   \ \ $(j_x)_{\gamma_1,1_{\nu}}$ \ \ &  \ \ $(j_x)_{\gamma_2,1_{\nu}}$ \ \ &  \ \ $J_0$ \ \ &
   \ \ $J_1$ \ \ \\ \hline
  1.395 & 1.530 & 0.106 & 1.167 & 0.778 & 0.525 & 0.625 & 36.37 & 127.49 \\ \hline
\end{tabular}
\end{center}

\noindent
$\omega_K$-- is the bandhead energy; \\
$(j_x)_{K,K'}$-- is the matrix element at Coriolis mixture;\\
$J_0$ and $J_1$-- is the inertia moment.

\section{Electric quadrupole transitions}

The reduced probability $E2$  transitions between type (3) states are given
as \cite{Bohr,Grigorev}:
\begin{eqnarray}
B(E2;I_iK_i\rightarrow I_fK_f)=\frac{1}{2I_i+1}\mid<I_fK_f\parallel
\widehat{m}(E2;\mu)\parallel I_iK_i>\mid^2
\end{eqnarray}

The calculated electric $E2$-- transitions from the state $I_iK_i$ to the
ground state bands $I_fK_f$ \cite{Usmanov,Mikhailov} are
\begin{eqnarray}
&&B(E2;I_iK_i\rightarrow I_fgr)=\{\sqrt{\frac{5}{16\pi}}[\Psi_{gr,gr}^{I_f}
\Psi_{gr,K_I}^{I_i}C_{I_i0;20}^{I_f0}+\sum_{n}\Psi_{K_n,gr}^{I_f}
\Psi_{K_n,K_i}^{I_i}C_{I_iK_n;20}^{I_fK_n}]+  \nonumber \\
&&\sqrt{2}[\Psi_{gr,gr}^{I_f}\sum_{n}\frac{(-1)^{K_n}m_{K_n}
\Psi_{K_n,K_i}^{I_i}}{\sqrt{1+\delta_{K_n,0}}}C_{I_iK_n;2-K_n}^{I_f0}
+\Psi_{gr,K_i}^{I_i}\sum_{n}\frac{m_{K_n}\Psi_{K_n,gr}^{I_f}}
{\sqrt{1+\delta_{K_n,0}}}C_{I_i0;2K_n}^{I_fK_n}]\}^2
\end{eqnarray}

\noindent
Here $m_K=<gr|\widehat{m}(E2)K^{\pi}>$, $K^{\pi}=0^+$, $2^+$ and $1_{\nu}^+$
are some constants, which can be obtained from the experimental data, $Q_0$
intrinsic quadrupole moment of nuclei, $C_{I_iK_i;2(K_i+K_f)}^{I_fK_f}$-- are
Klebsh - Gordan coefficients.

In the adiabatic approximation, the reduced probability of $E2$ transitions from
the $\beta$-- and $\gamma$-- vibrational bands are given by:
\begin{eqnarray}
B^{rot}(E2;I_i\beta\rightarrow I_fgr)=\mid m_0C_{I_i0;20}^{I_f0}\mid^2
\end{eqnarray}
\begin{eqnarray}
B^{rot}(E2;I_i\gamma\rightarrow I_fgr)=\mid m_2C_{I_i2;2-2}^{I_f0}\mid^2
\end{eqnarray}

\noindent
which allows $m_0$ and $m_2$, parameters calculation by using experimental
data \cite{Bekzhanov}. However, when they are close to each other, they strongly
mix even at $I=2$ and adiabatic approximation (9) and (10) cannot be used
(becomes inapplicable).

Calculated probabilities $E2$-- transitions from $0_n^+$ and $2_{n}^{+}$-- vibrating
bands onto levels of the ground band and resulted matrix elements $E2$-- transitions
from states $0_2^+$, $2_1^+$ and inside of the ground band are calculated.
Comparison of experimental data \cite{Bekzhanov,Suber} and calculated values of
quadrupole transitions probabilities are presented in Table 2. The calculated and
experimental \cite{Fahlander} values of the reduced matrix elements $E2$-- transitions
from states $0_2^+$, $2_1^+$ band to inside of the ground ($gr$) bands are given in
Table 3. One can note that $E2$-- calculations probabilities were calculated by using
the following values of $m_Š$ parameters:
$$m_{\beta_1}=12.65 e\cdot fm^2; \ \ m_{\beta_2}=0.374 e\cdot fm^2;$$
$$m_{\gamma_1}=1.5 e\cdot fm^2; \ \ m_{\gamma_2}=1.5 e\cdot fm^2; $$
$$m_{1_1}=14.0 e\cdot fm^2; \ \ m_{1_2}=9.0 e\cdot fm^2;$$

\noindent
The values of intrinsic quadrupole moment is taken as $Q_0=791 fm^2$ \cite{Usmanov}.
One can see from Tables 2 and 3 that the experimental data is well reproduced when
the above model is used.

\newpage
\textbf{Table 2. \ \ Reduced probability $E2$-- transitions }

\begin{tabular}{|c|c|c|c|c|c|} \hline
   \ $I_iK_i\rightarrow I_fK_f$ \ & \ \ experiment \ \ & \ \ theor \ \ &
   \ $I_iK_i\rightarrow I_fK_f$ \ & \ \ experiment \ \ & \ \ theor \ \ \\ \hline
    & $B(E2)$\cite{Bekzhanov,Suber} & $B(E2)$ &   &
   $B(E2)$\cite{Bekzhanov,Suber} & $B(E2)$ \\ \hline
$0\beta_1\rightarrow 0gr$ & -- & -- &
$2\gamma_1\rightarrow 0gr$ & 75(6) & 54 \\ \hline
$0\beta_1\rightarrow 2gr$ & 160(50) & 160 &
$2\gamma_1\rightarrow 2gr$ & 121(12) & 90 \\ \hline
$2\beta_1\rightarrow 2gr$ & 52(8) & 41 &
$4\gamma_1\rightarrow 4gr$ & 67(12) & 85 \\ \hline
$4\beta_1\rightarrow 4gr$ & -- & 29 &
$2\gamma_1\rightarrow 4gr$ & 6.8(7) & 5 \\ \hline
$2\beta_1\rightarrow 0gr$ & 14(1) & 31 &
$3\gamma_1\rightarrow 2gr$ & 152(11) & 143 \\ \hline
$2\beta_1\rightarrow 4gr$ & 140(20) & 90 &
$3\gamma_1\rightarrow 4gr$ & 79(6) & 64 \\ \hline
$0\beta_2\rightarrow 0gr$ & -- & -- &
$2\gamma_2\rightarrow 0gr$ & 32.(4) & 27 \\ \hline
$0\beta_2\rightarrow 2gr$ & 0.14 & 0.14 &
$2\gamma_2\rightarrow 2gr$ & 51(7) & 46 \\ \hline
$2\beta_2\rightarrow 2gr$ & 12(1) & 29 &
$4\gamma_2\rightarrow 4gr$ & -- & 47 \\ \hline
$4\beta_2\rightarrow 4gr$ & -- & 75 &
$2\gamma_2\rightarrow 4gr$ & 3.3(4) & 2 \\ \hline
$2\beta_2\rightarrow 0gr$ & 3.4(2) & 21 &
$3\gamma_2\rightarrow 2gr$ & 54(7) & 42 \\ \hline
$2\beta_2\rightarrow 4gr$ & 1.0(1) & 1 &
$3\gamma_2\rightarrow 4gr$ & 22(3) & 22 \\ \hline
\end{tabular}  \\

\textbf{Table 3. \ \ Reduced matrix elements E2 - transitions (eb) }

\begin{tabular}{|c|c|c|c|c|c|}  \hline
   \ $I_iK_i\rightarrow I_fK_f$ \ & \ \ Expe \ \cite{Fahlander} \ \ \ \ & \ \ Theor \ \ &
   \ $I_iK_i\rightarrow I_fK_f$ \ & \ \ Expe \ \cite{Fahlander} \ \ \ \ & \ \ Theor \ \ \\ \hline
  $2\gamma_1\rightarrow 0gr$ & 0.208$_{-0.040}^{+0.010}$ & 0.164 &
  $2gr\rightarrow 0gr$ & 2.45$\pm$0.12 & 2.50 \\ \hline
  $2\gamma_1\rightarrow 2gr$ & 0.250$_{-0.018}^{+0.016}$ & 0.212 &
  $4gr\rightarrow 2gr$ & 3.76$\pm$0.19 & 4.0 \\ \hline
  $2\gamma_1\rightarrow 4gr$ & 0.063$_{-0.004}^{+0.009}$ & 0.050 &
  $6gr\rightarrow 4gr$ & 5.34$\pm$0.27 & 5.05 \\ \hline
  $4\gamma_1\rightarrow 4gr$ & 0.46$_{-0.13}^{+0.08}$ & 0.260 &
  $8gr\rightarrow 6gr$ & 5.90$\pm$0.30 & 5.91 \\ \hline
  $4\gamma_1\rightarrow 2gr$ & 0.22$_{-0.05}^{+0.07}$ & 0.112 &
  $10gr\rightarrow 8gr$ & 6.71$\pm$0.34 & 6.66 \\ \hline
  $3\gamma_1\rightarrow 2gr$ & 0.326(11) & 0.316 &
  $12gr\rightarrow 10gr$ & 7.01$\pm$0.35 & 7.33 \\ \hline
  $3\gamma_1\rightarrow 4gr$ & 0.235(6) & 0.212 &
  $14gr\rightarrow 12gr$ & 8.12$_{-0.43}^{+0.63}$ & 7.95 \\ \hline
  $2\beta_1\rightarrow 0gr$ & 0.090$_{-0.009}^{+0.009}$ & 0.124 &
  $2\beta_1\rightarrow 4gr$ & 0.27$_{-0.08}^{+0.02}$ & 0.212 \\ \hline
  $2\beta_1\rightarrow 2gr$ & 0.162$_{-0.008}^{+0.071}$ & 0.143 &
  $4\beta_1\rightarrow 4gr$ & -- & 0.162 \\ \hline
\end{tabular}

\section{Electric Monopole transitions }

Electromagnetic transitions are important for in the understand the nature and
analyze various modes of nuclear excitations. Type of electric monopole
transitions are of special importance. They are connected with interesting
and poorly studied aspects of nuclear structure.

The main of $E0$-- transition is due to Columb interaction of nucleon with atomic
electrons. The other, ones as well as electromagnetic and other interactions are
not important. One offent - phonon $E0$-- transitions are forbidden because of
angular momentum conservation.

The monopole transitions are usually accompanied by strong quadrupole transitions.
When it is impossible to determine the life - time of the level being studied,
one measures the magnitude of $E0/E2$-- mixing
\begin{eqnarray}
X_I(E0/E2)=\frac{B(E0;i\rightarrow f)}{B(E2;i\rightarrow f)}=
\frac{e^2R_0^4\rho(E0;i\rightarrow f)}{B(E2;i\rightarrow f)}
\end{eqnarray}

\noindent
where $\rho$-- is the dimensionless matrix element of $E0$-- transition and
$R_0=r_0A^{1/3}$ is nucleus charge radius. The transitions between monopole
states are compared with $E2$-- transitions from $0^+$-- level to the $2^+$--
levels, nearest to the last state in $E0$-- transition.

For the reduced of probabilities of $E0$-- transitions from the
$K^{\pi}=0_{\nu}^{+}$ state bands to the ground rotational bands within present
model we find that \cite{Usmanov}:
\begin{eqnarray}
X_I(E0;I0_i\rightarrow Igr)=\{\sum m_{0}^{'}(\Psi_{0_{\nu},0_2}^{I}\Psi_{gr,gr}^{I}+
\Psi_{gr,0_2}^{I}\Psi_{0,gr}^{I})\},
\end{eqnarray}

\noindent
where $\nu$-- quantum number of $0^+$-- bands, that is included to the basis
states of the Hamiltonian (2), $m_{0}^{'}=<gr\mid m(E0)\mid 0_{\nu}^+>$ are the
matrix elements between intrinsic wave functions for the ground - state and the
$0_{\nu}^+$-- bands, which are numerical parameters and determined by experimental
data. For the adiabatic case equation (11) may be rewritten as:
\begin{eqnarray}
X_I(E0/E2)=X_0(E0/E2)\frac{(2I-1)(2I+3)}{I(I+1)}
\end{eqnarray}
where
\begin{eqnarray}
X_0(E0/E2)=\frac{B(E0;00_{\nu}\rightarrow 0gr)}{B(E2;00_{\nu}\rightarrow 2gr)}=
[\frac{m_{0_{\nu}}^{'}}{m_{0_{\nu}}}]^2
\end{eqnarray}

\noindent
Using experimental data $m_{0_{\nu}}^{'}$ for the $0_{\nu}^+$ level, it is possible
to determine $X_0^{exp}(E0/E2)$  by (14).

In case when the parameter $m_{0_{\nu}}$ is obtained by adiabatic equation (9),
definition of parameters $m_{0_{\nu}}^{'}$ by (14) is accurate since $0_{\nu}^+$
levels are not pretreated by the Coriolis forces. The value of parameters of
$m_{0_{\nu}}^{'}$, obtained from the experimental data \cite{Bekzhanov}, are found to be
$m_{0_{2}}^{'}=2.154 e\cdot fm^2$ and  $m_{0_{3}}^{'}=0.641 e\cdot fm^2$.

The values of $X_I(E0/E2)$ and $\rho(E0)$ for the transitions from state
$\beta_1$--, $\beta_2$--, $\gamma_1$-- and $\gamma_2$-- bands are given in
Table 4, along with the experimental data \cite{Suber,Demidov}. The
experimentally observed $K$-- forbidden $E0$-- transitions may be described
within our phenomenological model. Availability of these transitions can be
explained by mixture of states $\gamma_1$-- and $\gamma_2$-- bands with
$\beta_1$--, $\beta_2$-- state bands.

\begin{center}
\textbf{Table 4. \ \ Characteristics of $E0$ - transition between the excitations states }

\begin{tabular}{|c|c|c|c|c|} \hline
   \ $IK_i\rightarrow IK_f$ \ \ &  \ \ experiment \ \ &  &  \ \ theor \ \ &   \\ \hline
    &  \ \ $X_I$ \cite{Suber,Demidov} \ \ &  \ \ $\rho(E0)$ \cite{Demidov} \ \ &
     \ \ $X_I$ \ \ &  \ \ $\rho(E0)$ \ \  \\ \hline
  $0\beta_1\rightarrow 0gr$ & 0.028(4) & 0.048(7) & 0.029 & 0.048 \\
  $2\beta_1\rightarrow 2gr$ & <0.106 & 0.038(6) & 0.11 & 0.11 \\
  $4\beta_1\rightarrow 4gr$ & 0.020 & 0.02(2) & 0.15 & 0.14 \\ \hline
  $2\gamma_1\rightarrow 2gr$ & <0.002 & 0.02(2) & 0.002 & 0.02 \\
  $4\gamma_1\rightarrow 4gr$ & 0.029(20) & 0.05(2) & 0.006 & 0.05 \\ \hline
\end{tabular}
\end{center}

\section{Magnetic dipole transitions}

Within the present model, the probability of $M1$-- transition from the
one - phonon state to the ground rotation bands levels is a given by
equation \cite{Usmanov}:
\begin{eqnarray}
B(M1;IK\rightarrow I'0_{gr})&=&\frac{3}{4\pi}\mid\sum_{K_1=1}^{2}
\Psi_{K_1,K}^{I}\Psi_{K_1,gr}^{I'}C_{IK_1;10}^{I'K_1}(g_{K'}-g_{R})+ \nonumber \\
&+&\frac{\sqrt{6}}{10}m_{1}^{'}\Psi_{gr,gr}^{I'}\Psi_{1,K}^{I}
C_{I1;10}^{I'0}\mid^2(\frac{e\hbar}{2Mc})^2.
\end{eqnarray}

\noindent
here $g_K$-- intrinsic $g$-- factor bands with $K\neq 0$.

The rare - land and transuranium region on deformation nuclei of $g_R$--
factor have the following value $g_R\cong 0.4\pm 0.1$ \ ($g_R=Z/A$) \cite{Bohr} from
systematic. In the adiabatic approximation, the transitions from
$IK^{\pi}=11^+$ state we have \cite{Usmanov}:
\begin{eqnarray}
B(M1;11^+\rightarrow 00_{gr}^+)=\frac{3}{4\pi}(\frac{e\hbar}{2Mc})^2
\cdot 0.02 (m_{1}^{'})^2.
\end{eqnarray}

Using the experimental data of $M1$-- transitions probability it is impossible
to calculate $m_{1}^{'}$ parameters. For the magnetic momentum of state we have
the following equation:
\begin{eqnarray}
\mu_{K}(I)=g_{R}I+\sum_{K'=1}\mid\Psi_{K',K}^{I}\mid^2(g_{K'}-g_{R})+
\frac{\sqrt{3}}{10}m_{1}^{'}\Psi_{gr,K}^{I}\Psi_{1,K}^{I}\sqrt{\frac{I}{I+1}}.
\end{eqnarray}

\noindent
Usually, instead of the $M1$-- transitions reduced probability one can use the
coefficient of the multipole mixture $\delta$, the value of which was determined
experimentally by
\begin{eqnarray}
\delta(I_iK_i\rightarrow I_fK_f)= 0.834\cdot E_{\gamma} (MeV)
\frac{<I_fK_f\parallel\widehat{m}(E2)\parallel I_iK_i>}
{<I_fK_f\parallel\widehat{m}(M1)\parallel I_iK_i>}
(\frac{e\cdot b}{\mu_{Ya}}).
\end{eqnarray}

In the adiabatic approximation for the transition inside one state band
with $K\neq 0$ the equation (18):
\begin{eqnarray}
\delta =0.933 \frac{eQ}{g_{K}-g_{R}}\frac{E_{\gamma}}{\sqrt{I^2-1}}.
\end{eqnarray}

\noindent
where, $E_{\gamma}$-- energy of $\gamma$-- transition. The theoretical and
experimental values of $\delta(E2/M1)$: \cite{Gresswell,Bengtsson} 
are given in Table 5.

\begin{center}
\textbf{Table 5. \ \ The multiple mixture coefficients $\delta(E2/M1)$  }

\begin{tabular}{|c|c|c|c|c|} \hline
   \ \ $I_iK_i\rightarrow I_fK_f$ \ \ &  \ \ \ $E_{\gamma}$ \ \ \ & \ \ \ \ \ \ exp.
  \cite{Gresswell,Kracikova,Singh,Demidov} \ \ \ \ \ \ &  \ \ theor. \cite{Kracikova} \ \ &  \ \ \ theor. \ \ \ \\ \hline
  $2\gamma_1\rightarrow 2gr$ & 1.387 & -5.1$_{-1.6}^{+1.1}$ & -4.42 & -3.5 \\ \hline
  $3\gamma_1\rightarrow 2gr$ & 1.471 & -7.3$_{-1.1}^{+0.8}$ & -6.01 & -6.5 \\ \hline
  $3\gamma_1\rightarrow 4gr$ & 1.289 & 2.8$_{-1.0}^{+0.7}$ & -- & -4.4 \\ \hline
  $4\gamma_1\rightarrow 4gr$ & 1.398 & -1.1$_{-0.5}^{+0.2}$ and -42.0$_{-\propto}^{+36.0}$ & -2.20 & -1.8 \\ \hline
  $5\gamma_1\rightarrow 4gr$ & 1.532 & -- & -- & -3.9 \\ \hline
  $5\gamma_1\rightarrow 6gr$ & 1.252 & -- & -- & -2.9 \\ \hline
  $2\gamma_2\rightarrow 2gr$ & 1.530 & 8.0$_{-1.5}^{+2.0}$ & -2.36 & -3.7 \\ \hline
  $3\gamma_2\rightarrow 2gr$ & 1.622 & 17.9$_{-2.2}^{+2.9}$ & -2.15 & -3.0 \\ \hline
  $3\gamma_2\rightarrow 4gr$ & 1.441 & 5.8$_{-0.5}^{+0.7}$ & -1.56 & -2.2 \\ \hline
  $4\gamma_2\rightarrow 4gr$ & 1.543 & 9.0$_{-3.0}^{+11.0}$ and -0.84$_{-0.11}^{+0.09}$ & -1.23 & -2.5 \\ \hline
  $5\gamma_2\rightarrow 4gr$ & 1.667 & 6.4$_{-0.7}^{+1.0}$ & -1.2 & -1.6 \\ \hline
  $2\beta_1\rightarrow 2gr$ & 1.039 & 5.0$_{-1.6}^{+2.5}$ & 5.80 & 1.7 \\ \hline
  $4\beta_1\rightarrow 4gr$ & 1.026 & 0.87$_{-0.13}^{+0.13}$ & 2.89 & 0.78 \\ \hline
  $6\beta_1\rightarrow 6gr$ & 0.997 & 0.63$_{-0.07}^{+0.07}$ & 1.82 & 0.46 \\ \hline
  $2\beta_2\rightarrow 2gr$ & 1.398 & $\geq$ 2.2 & -0.50 & 3.5 \\ \hline
\end{tabular}
\end{center}

From Table 5, one can see, that the used model \cite{Usmanov} with the
two $M1$-- transitions operator parameters reproduces well multipole mixture
coefficients $\delta(E2/M1)$ for $\beta_1\rightarrow gr$, $\gamma_1\rightarrow gr$ and
$\gamma_2\rightarrow gr$ transitions. It should be noted that the multipole
mixture coefficients of $\delta(E2/M1)$ for the $\beta_1\rightarrow gr$,
$\gamma_1\rightarrow gr$ and $\gamma_2\rightarrow gr$ transitions are equal
to zero in the adiabatic approximation.

\section{Conclusions}

Our calculations taking into account the Coriolis mixture of positive - parity
states give good agreement with experimental data.

The energy spectrum experimental data, probabilities of electric quadrupole and
monopole transitions from $\beta_1$--, $\beta_2$--, $\gamma_1$-- and $\gamma_2$--
band to the rotational levels ground state bands are in good agreement in the
present model. Thus within our theoretical analysis it is possible to explain the
$E0$-- transition from the state $\gamma_1$-- and $\gamma_2$-- band and $M1$--
transition from the state $\beta_1$--, $\beta_2$--, $\gamma_1$-- and $\gamma_2$--
bands to the ground bands level, which is forbidden by adiabatic approximation.


\begin{thebibliography}{99}

\bibitem{Bohr} A. Bohr and B.R. Mottelson, "Nuclear Structure Vol. {\bf 1.2}"
// (Benjamin, New York, 1975), Mir. Moscow (1971), (1977).
\bibitem{Grigorev} E.P. Grigorev, V.G. Solovev "Structure Deformations nuclear"
// M. Nauk, (1974).
\bibitem{Bekzhanov} R.B. Bekzhanov, V.M. Belinkiy, et al. "The directory on the
nuclear physics ". // Tashkent, FAN, Vol.{\bf 1,2}, (1989).
\bibitem{Fahlander} C.Fahlander, B.Varnestig, F.Backlin et al. // Nucl.Phys.
A{\bf 541}, 157 (1992).
\bibitem{Youhana} H.M.Youhana, R.Al-Obeidi, V.F.Al-Amili et al. // Nucl.Phys.
A{\bf 458}, p.51 (1986).
\bibitem{Usmanov} Ph.N.Usmanov, I.N.Mikhailov.// Fiz.Elem.Chastits At.
Yadra Vol.{\bf 28}, p.887 (1997).
\bibitem{Gresswell} J.R.Gresswell, P.D.Forsyth, D.G.E.Martin, R.Morgan // J.Phys.
G{\bf 7}, p.235-261, (1981).
\bibitem{Suber} A.R.Suber et al. // J. Phys. G{\bf 14}, p.87 (1988).
\bibitem{Voinonova} N.A.Voinonova-Eliseeva and I.A. Mitropol'skaya //
Fiz.Elem.Chastits At. Yadra Vol.{\bf 17(6)}, p.1173 (1986).
\bibitem{Harris} S.M.Harris // Phys. Rev. B{\bf 138}, p.509, (1965).
\bibitem{Bengtsson} R.Bengtsson, S.Frauendorf // Nucl.Phys. A{\bf 327}, 139 (1979).
\bibitem{Mikhailov} I.N.Mikhailov, Ph.N.Usmanov, A.A.Okhunopv et al. //
Izv. Akad. Nauk RAN, Ser. Fiz. Vol.{\bf 57(1)}, p.17 (1993).
\bibitem{Kracikova} T.I.Kracikova, S.Davaa, M.Finger et al. // J.Phys. G{\bf 10}, p.1115 (1983).
\bibitem{Singh} B.Singh // Nucl. Data Sheets, Vol.{\bf 75}, p.199 (1995).
\bibitem{Demidov} A.M.Demidov, L.I.Covor, V.A.Kurkin, I.V.Mikhailov //
Nucl.Phys., Vol.{\bf 69(7)}, p.1205 (2006).
\end{thebibliography}
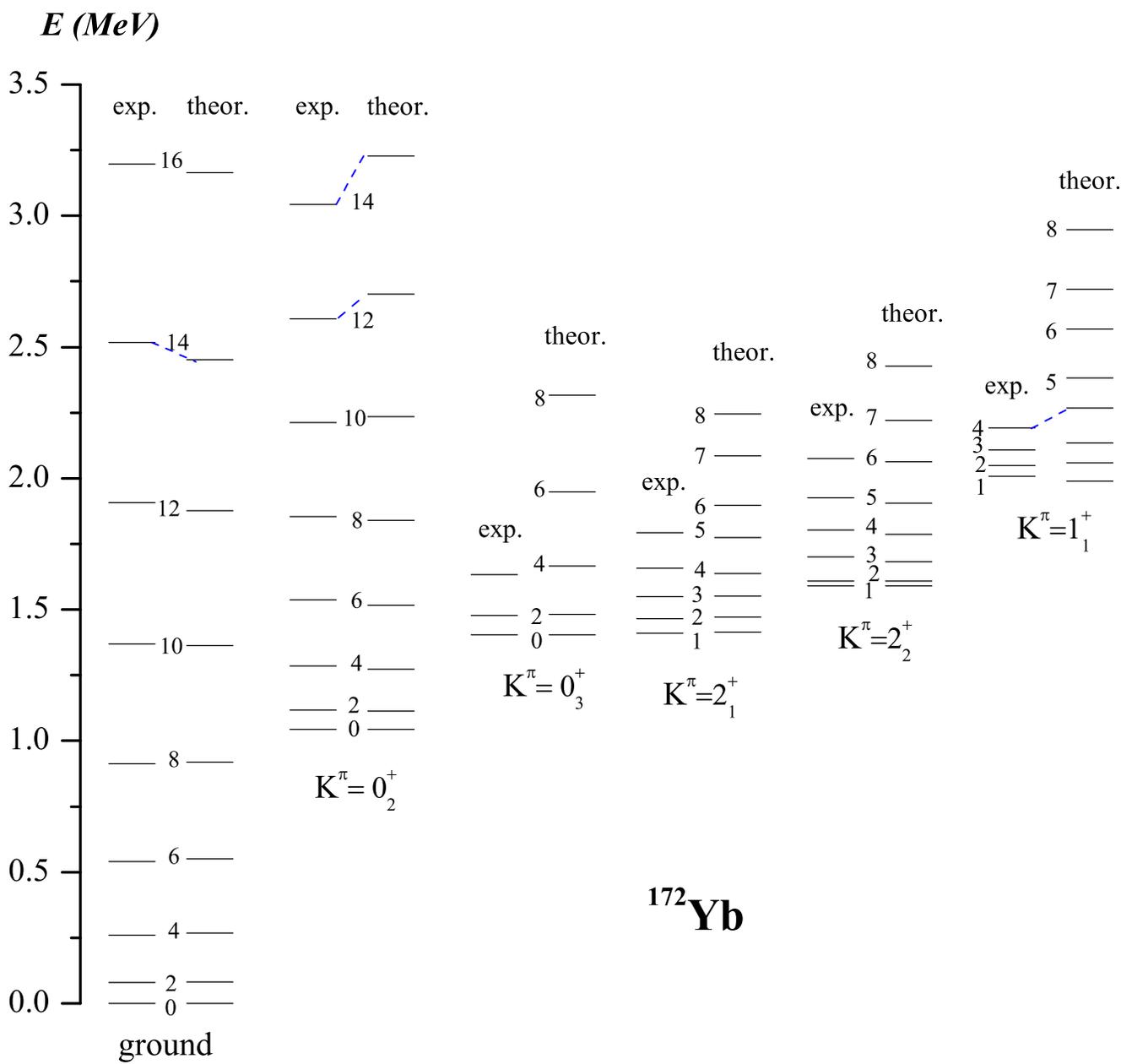
\end{document}